\documentclass[12pt]{article}

\newcommand{\be}{\begin{equation}}
\newcommand{\ee}{\end{equation}}
\newcommand{\bi}[1]{\vspace{-3mm} \bibitem{#1}}

\usepackage{epsfig,amsmath}
\usepackage{graphics}

\begin{document}

\begin{center}
{\Large \bf Nonholonomic Constraints with Fractional Derivatives}
\vskip 5 mm

{\large \bf  Vasily E. Tarasov$^{1,2}$ and George M. Zaslavsky$^{2,3}$}\\

\vskip 3mm

{\it $1)$ Skobeltsyn Institute of Nuclear Physics, \\
Moscow State University, Moscow 119992, Russia } \\
{E-mail: tarasov@theory.sinp.msu.ru} \\
{\it $2)$ Courant Institute of Mathematical Sciences, New York University \\
251 Mercer St., New York, NY 10012, USA, and }\\ 
{\it $3)$ Department of Physics, New York University, \\
2-4 Washington Place, New York, NY 10003, USA } \\
\end{center}

\begin{abstract}
We consider the fractional generalization of nonholonomic constraints
defined by equations with fractional derivatives and
provide some examples.
The corresponding equations of motion 
are derived using variational principle.
\end{abstract}






\section{Introduction}


The theory of derivatives of non-integer order goes back 
to Leibniz, Liouville, Riemann, Grunwald, and Letnikov 
\cite{SKM,OS,Podlubny}. Fractional analysis proved to be useful 
in mechanics and physics.
In a fairly short time the list of such 
applications continuously grows.
The applications include 
chaotic dynamics \cite{Zaslavsky1,Zaslavsky2},
material sciences \cite{Hilfer,C2,Nig1,Nig4}, 
mechanics of fractal and complex media 
\cite{Mainardi,Media,Physica2005},
quantum mechanics \cite{Laskin,Naber}, 
physical kinetics \cite{Zaslavsky1,Zaslavsky7,SZ,ZE},
plasma physics \cite{CLZ,Plasma2005}, 
electromagnetic theory \cite{Lutzen,Mil2,Plasma2005},  
long-range dissipation \cite{GM}, 
non-Hamiltonian mechanics \cite{nonHam,FracHam},
long-range interaction \cite{Lask,TZ3}, 
anomalous diffusion and transport theory 
\cite{Zaslavsky1,Montr,Uch,MK}.

In this paper, we consider the fractional generalization of 
nonholonomic constraints such that the constraint equations 
consist of fractional derivatives, called fractional constraints. 
The corresponding equations of motion will be 
derived by the d'Alembert-Lagrange principle 
and some simple examples are considered. 

In Sec 2, we provide a brief review of nonholonomic systems,  
fix notations and convenient references. In Sec. 3,
we consider fractional generalizations of nonholonomic constraints.
Some examples are considered. In Sec. 4, we discuss the applicability
of the stationary action principle for fractional constraints.
Finally, a short conclusion is given in Sec. 5.

\section{Nonholonomic Constraints with Integer Derivatives} 

In this section, a brief review of 
nonholonomic systems is considered to 
fix notations and provide convenient references
\cite{Dob}.

\subsection{Lagrange Equations for Nonholonomic System}

It is known that the d'Alembert-Lagrange principle
allows us to derive equations of motion with holonomic and
nonholonomic constraints. For N-particle system
it has the form of the variation equation
\be \label{con2}
\left(\frac{d(m {\bf v}_i)}{dt} -
{\bf F}_i \right)\delta {\bf r}_i=0, \ee
where ${\bf r}_i$ ($i=1...N$) is a radius-vector of $i$th particle,
${\bf v}_i=\dot{\bf r}_i$ is a velocity,
and ${\bf F}_i$ is a force that acts on $i$th particle, 
and the sum over repeated index $i$ is from 1 to $N$.
To exclude holonomic constraints, the general coordinates $q^k$ 
($k=1,...n$) are used. Here, $n=3N-m$ is a number of degrees of freedom, 
where $m$ is a number of holonomic constraints.
Then ${\bf r}_i$ is a function of generalized coordinates and time:
${\bf r}_i={\bf r}_i(q,t)$.
Using $\delta {\bf r}_i=(\partial {\bf r}_i/ \partial q^k) \delta q^k$,
Eq. (\ref{con2}) gives
\be \label{dl} \left(\frac{d(m {\bf v}_i)}{dt}
\frac{\partial {\bf r}_i}{\partial q^k}-
{\bf F}_i \frac{\partial {\bf r}_i}{\partial q^k} \right)
\delta q^k=0 , \ee
and sum over repeated index $k$ is from 1 to $n$. 
Then, we define \cite{Pars} the generalized forces:
\[ Q_k={\bf F}_i \frac{\partial {\bf r}_i}{\partial q^k} \quad k=1,...,n. \]
By usual transformations \cite{Pars}
\[ \frac{d(m {\bf v}_i)}{dt}
\frac{\partial {\bf r}_i}{\partial q^k}
=\frac{d}{dt}\left(m{\bf v}_i
\frac{\partial {\bf r}_i}{\partial q^k}\right)-
(m{\bf v}_i)\frac{d}{dt}
\frac{\partial {\bf r}_i}{\partial q^k}=\]

\[ =\frac{d}{dt}
\left(m{\bf v}_i
\frac{\partial {\bf v}_i}{\partial \dot{q}^k}\right)-
m{\bf v}_i\frac{\partial {\bf v}_i}{\partial q^k}=\]
\[ =\frac{d}{dt}
\frac{\partial}{\partial \dot{q}^k}\left(\frac{m}{2}
{\bf v}_i{\bf v}_i\right)-\frac{\partial}{\partial q^k}
\left(\frac{m}{2} {\bf v}_i{\bf v}_i\right)
=\frac{d}{dt}\frac{\partial T}{\partial \dot{q}^k}
-\frac{\partial T}{\partial q^k}, \]
we transform Eq. (\ref{dl}) into
\be \label{Jor} \left(\frac{d}{dt}
\frac{\partial T}{\partial \dot{q}^k}-\frac{\partial T}{\partial q^k}
-Q_k \right) \delta q^k=0 , \ee
where $T={m {\bf v}^2}/{2}$ is a kinetic energy. Using
\be
{\bf v}_i=\frac{d{\bf r}_i(q,t)}{dt}=
\frac{\partial {\bf r}_i}{\partial q^k} \frac{dq^k}{dt}+
\frac{\partial {\bf r}_i}{\partial t} , 
\ee
we get
\[ T=\frac{m}{2}\left(g_{kl}(q,t) \dot{q}^k \dot{q}^l+
2g_{k}(q,t)\dot{q}^k +g(q,t) \right) , \]
where
\be
g_{kl}(q,t)=\frac{\partial {\bf r}_i}{\partial q^k}
\frac{\partial {\bf r}_i}{\partial q^l}, \quad
g_{k}(q,t)=\frac{\partial {\bf r}_i}{\partial q^k}
\frac{\partial {\bf r}_i}{\partial t}, \quad
g(q,t)=\frac{\partial {\bf r}_i}{\partial t}
\frac{\partial {\bf r}_i}{\partial t}.
\ee

For the nonholonomic constraint, 
\be \label{Rk1} R_k \delta q^k=0 , \ee
where $R_k$ is a reaction force of the constraint
\be \label{C} f(q,\dot{q})=0 , \ee
and the variations $\delta q^k$
are defined \cite{Tchet,Dob} by 
\be \label{Ch} \frac{\partial f}{\partial \dot{q}^k} \delta q^k=0 . \ee
Equation (\ref{Ch}) is called Chetaev's condition \cite{Dob}.
Comparing  Eqs. (\ref{Rk1}) and (\ref{Ch}), we obtain
\be \label{Rlf}
R_k=\lambda \frac{\partial  f}{\partial \dot{q}^k}, \ee
where $\lambda$ is a Lagrange multiplier.
Chetaev's definition of variations states that the actual constrained 
motion should occur along a trajectory obtained by normal projection
of a force onto a constraint hypersurface. 
The constraint force $R_k$ is minimum when $R_k$ is chosen 
perpendicular to the constraint surface or parallel 
to the gradient $\partial f/ \partial \dot{q}_k$.

In general, the nonholonomic system is subjected to action of 
the generalized force $Q_k$ and the constraint force $R_k$.
Then the variational equation is
\be \label{Jor2} \left(\frac{d}{dt}
\frac{\partial T}{\partial \dot{q}^k}-\frac{\partial T}{\partial q^k}
-Q_k -R_k \right) \delta q^k=0 . \ee
From (\ref{Rlf}), we obtain 
\be \label{Jor2b} \left(\frac{d}{dt}
\frac{\partial T}{\partial \dot{q}^k}-\frac{\partial T}{\partial q^k}
-Q_k -\lambda \frac{\partial  f}{\partial \dot{q}^k}\right)
\delta q^k=0 .\ee
In Eq. (\ref{Ch}), we can consider $\delta \dot{q}^s$,
$s=1,2,...,n-1$, as independent variations.
Then $\delta \dot{q}^n$ is not independent, and Eq. (\ref{Ch}) gives
\[ \delta q^n=-\left( \frac{\partial f}{\partial \dot{q}^n} \right)^{-1}
\sum^{n-1}_{s=1}\frac{\partial f}{\partial \dot{q}^s} \delta q^s . \]
Suppose that $\lambda$ satisfies the equation
\be \frac{d}{dt}
\frac{\partial T}{\partial \dot{q}^n}-\frac{\partial T}{\partial q^n}
-Q_n -\lambda \frac{\partial  f}{\partial \dot{q}^n}=0 . \ee
Then the term with $k=n$ in (\ref{Jor2b}) is equal to zero,
and Eq. (\ref{Jor2b}) has $n-1$ terms with $k=1,...,n-1$.
In Eq. (\ref{Jor2b}), the variations with $k=1,2,...,n-1$ 
are independent, and the sum is separated  on $n-1$ equations. 
As the result, Eq. (\ref{Jor2b}) is equivalent to 
\be \label{TT} \frac{d}{dt} \frac{\partial T}{\partial \dot{q}^k}
-\frac{\partial T}{\partial q^k}=
Q_k +\lambda\frac{\partial f }{\partial \dot{q}^k} , \quad (k=1,...,n). \ee
Equations (\ref{C}) and (\ref{TT}) form a system of $n+1$ equations
with $n+1$ unknowns $\lambda$ and $q^{k}$, where $k=1,...,n$.
Solutions of these equations describe particles motion
as a motion of system with
nonlinear nonholonomic constraint (\ref{C}).

\subsection{Nonholonomic System as a Holonomic One}

In this section, we present
the equations of motion with nonholonomic constraint
as equations for a holonomic system.

The canonical momenta $p^k$ are defined by 
\be \label{mom}
p_k=\frac{\partial T}{\partial \dot{q}^k}=
mg_{kl}(q,t)\dot{q}^{l}+m g_{k}(q,t), \quad (k=1,...,n) . \ee
Using (\ref{mom}), we can define
\be \label{tf} \tilde f(p,q,t)=f(\dot{q}(q,p,t),q,t). \ee
Suppose that the constraint is integral of motion.
Then the total time derivative of (\ref{tf}) gives
\be \label{ttd} \frac{d \tilde f}{dt}=0, \quad
\frac{\partial \tilde f}{\partial p_k} \dot{p}_k+
\frac{\partial \tilde f}{\partial q^k}\dot{q}^k+
\frac{\partial \tilde f}{\partial t}=0. \ee
Substitution of (\ref{TT}) into (\ref{ttd}) get
\be \label{co1}
\frac{\partial \tilde f}{\partial p_k}
\left(\frac{\partial T}{\partial q^k}+Q_k+
\lambda \frac{\partial \tilde f}{\partial \dot{q}^k}  \right)+
\frac{\partial \tilde f}{\partial q^k}\dot{q}^k+
\frac{\partial \tilde f}{\partial t}=0 . \ee
From Eq. (\ref{co1}), we obtain 
\be \label{LM} \lambda 
=-\left(\frac{\partial \tilde f}{\partial p_k}
\frac{\partial \tilde f}{\partial \dot{q}^k}\right)^{-1}
\left( \frac{\partial \tilde f}{\partial p_k}
\left(\frac{\partial T}{\partial q^k}+Q_k \right)+
\frac{\partial \tilde f}{\partial q^k}\dot{q}^k+
\frac{\partial \tilde f}{\partial t} \right) . \ee
Then the Lagrange equations (\ref{TT}) have the form
\be \label{LE2b}
\frac{d}{dt} \frac{\partial T}{\partial \dot{q}^k}-
\frac{\partial T}{\partial q^k}=Q_k-
\frac{\partial  \tilde f}{\partial \dot{q}^k}
\left(\frac{\partial \tilde f}{\partial p_k}
\frac{\partial \tilde f}{\partial \dot{q}^k}\right)^{-1}
\left( \frac{\partial \tilde f}{\partial p_k}
\left(\frac{\partial T}{\partial q^k}+Q_k \right)+
\frac{\partial \tilde f}{\partial q^k}\dot{q}^k+
\frac{\partial \tilde f}{\partial t} \right) . \ee
Equations (\ref{LE2b}) describes the motion of a holonomic
system with $n$ degrees of freedom. For any trajectory of the 
system in the phase space, we have $\tilde f=0$.
If the initial values $q_k(0)$ and $\dot{q}_k(0)$ satisfy the
constraint condition $f(q(0),\dot{q}(0),t_0)=0$,
then the solution of Eq. (\ref{LE2b}) is a motion
of the nonholonomic system.

Let us define a generalized force $\Lambda_k=Q_k+R_k$, which
depends on generalized velocities $\dot{q}^k$,
generalized coordinates $q^k$, and time $t$.
If  
\[ \frac{\partial \Lambda_k}{\partial \dot{q}^m}+
\frac{\partial \Lambda_m}{\partial \dot{q}^k}=0, \]
\[ \frac{\partial \Lambda_k}{\partial q^m}+
\frac{\partial \Lambda_m}{\partial q^k}=
\frac{1}{2}\frac{d}{dt}\left(
\frac{\partial \Lambda_k}{\partial \dot{q}^m}-
\frac{\partial \Lambda_m}{\partial \dot{q}^k} \right), \]
known as the Helmholtz conditions,
are satisfied, then a generalized potential $U=U(\dot{q},q,t)$ exists and
\[ \frac{d}{dt} \frac{\partial U}{\partial \dot{q}^k}-
\frac{\partial U}{\partial q^k}=\Lambda_k.  \]
In this case, the Hamilton variational principle has 
the form of the stationary action principle. 
In order to use this principle for a nonholonomic system, we should
consider such trajectories that their initial conditions 
satisfy the constraint equation (\ref{C}).

Note that nonholonomic constraint (\ref{C}) and
non-potential generalized force $Q_k$ can be compensated such that
resulting generalized force $\Lambda_k$ is a generalized potential force,
and system is a Lagrangian and non-dissipative system 
with holonomic constraints.

\section{Constraints with Fractional Derivatives}

\subsection{Fractional Derivatives}

The fractional derivative has different definitions \cite{SKM,OS}, 
and exploiting any of them depends on the kind of 
the problems, initial (boundary) conditions, and 
the specifics of the considered physical processes.
The classical definition is the so-called Riemann-Liouville
derivative \cite{SKM,OS}
\[
_a{\cal D}^{\alpha}_{t}f(x)=
\frac{1}{\Gamma(m-\alpha)} \frac{\partial^m}{\partial x^m}
\int^{x}_{a} \frac{f(z) dz}{(x-z)^{\alpha-m+1}},
\]
\be \label{D2}
_t{\cal D}^{\alpha}_{b}f(x)=
\frac{(-1)^m}{\Gamma(m-\alpha)} \frac{\partial^m}{\partial x^m}
\int^{b}_x \frac{f(z) dz}{(z-x)^{\alpha-m+1}} ,
\ee
where $m-1<\alpha<m$. Due to reasons, concerning the initial conditions,
it is more convenient to use the Caputo fractional derivatives
\cite{C2,C1,C3}.
Its main advantage is that the initial conditions take the same
form as for integer-order differential equations. \\

{\bf Definition.}
{\it The Caputo fractional derivatives are defined by the equations
\be \label{1}
_aD^{\alpha}_tf(t)=\frac{1}{\Gamma(m-\alpha)} 
\int^t_a \frac{f^{(m)}(\tau)}{(t-\tau)^{\alpha+1-m}} d \tau ,
\ee
\be \label{2}
_tD^{\alpha}_b f(t)=\frac{1}{\Gamma(m-\alpha)} 
\int^b_t \frac{f^{(m)}(\tau)}{(t-\tau)^{\alpha+1-m}} d \tau ,
\ee
where $m-1 < \alpha < m$, and $f^{(m)}(\tau)=d^m f(\tau)/d\tau^m$. } \\

{\bf Proposition 1.} 
{\it The total time derivative of the Caputo fractional derivative of 
order $\alpha$ can be presented as a fractional 
derivative of order $\alpha+1$ by
\be \label{3}
\frac{d}{dt} \ _aD^{\alpha}_t f(t)= _aD^{\alpha+1}_t f(t)+
\frac{t^{m-\alpha-1}}{\Gamma(m-\alpha)} f^{(m)}(a) ,
\ee
where $m=[\alpha]+1$, and $[...]$  means floor function.}\\

{\bf Proof}. 
The definition (\ref{1}) can be presented in the form
\be \label{4}
_aD^{\alpha}_{t} f= \ _aJ^{m-\alpha}_{t} D^{m}_t f ,\ee
where $ \ _aJ^{m-\alpha}_{t}$ is a fractional integral
\be
_aJ^{\varepsilon}_tf(t)=\frac{1}{\Gamma(\varepsilon)} 
\int^t_a \frac{f(\tau)}{(t-\tau)^{1-\varepsilon}} d \tau .
\ee
The operations $D^1_t$ and $ J^{\varepsilon}_t$ do not commute:
\be \label{comm}
D^1_t \ _aJ^{\varepsilon}_t f(t)= _aJ^{\varepsilon}_t D^1_t f(t)+
\frac{t^{\varepsilon-1}}{\Gamma(\varepsilon)} f(a).
\ee
From (\ref{4}), we have
\be \label{24}  
\frac{d}{dt} \ _aD^{\alpha}_{t} f(t)=D^1_t \ _aJ^{\varepsilon}_{t} D^m_t f(t) 
=D^1_t \ _aJ^{\varepsilon}_t f^{(m)}(t),
\ee
where $ _aJ^{\varepsilon}_t $ is a 
fractional integration of order $\varepsilon=m-\alpha$.
Using (\ref{comm}) and (\ref{24}), we get
\[ 
D^1_t J^{\varepsilon}_t D^m_t f(t) 
= _aJ^{\varepsilon}_t D^1_t f^{(m)}(t)+
\frac{t^{\varepsilon-1}}{\Gamma(\varepsilon)} f^{(m)}(a)= \]
\be \label{27b} = \ _ aJ^{\varepsilon}_t f^{(m+1)}(t)+
\frac{t^{\varepsilon-1}}{\Gamma(\varepsilon)} f^{(m)}(a)= 
\ _aD^{\alpha+1}_t f(t)+
\frac{t^{\varepsilon-1}}{\Gamma(\varepsilon)} f^{(m)}(a) .
\ee
Substitution of (\ref{27b}) into (\ref{24}) proves (\ref{3}). 


\subsection{Fractional Equations of Motion}

Asume that the constraint equation has fractional derivatives:
\be \label{fnhc}
f(q,\dot{q},\ _aD^{\alpha}_t q,\ _tD^{\alpha}_b q)=0, \ee
i.e. it is a fractional differential equation
\cite{Podlubny}.
Such constraint can be called fractional nonholonomic constraint. 
Since Eq. (\ref{fnhc}) has also derivatives of integer order, we can use
the Chetaev definition of variation (\ref{Ch}) and
the Lagrange equations (\ref{TT}).
For generalized potential forces 
\[ Q_k=\frac{d}{dt} \frac{\partial U}{\partial \dot{q}_k} -
\frac{\partial U}{\partial q_k} ,\quad (k=1,...,n) , \]
and we can rewrite Eq. (\ref{TT}) as
\be \label{5} \frac{d}{dt} \frac{\partial L}{\partial \dot{q}_k} -
\frac{\partial L}{\partial q_k}=
\lambda \frac{\partial f}{\partial \dot{q}_k},\quad (k=1,...,n) , \ee
where $L=T(q,\dot{q})-U(q,\dot{q})$ is the Lagrangian.
To simplify our calculations, we consider
\be \label{L}
L=L(q,\dot{q})=\frac{1}{2} \sum^n_{k=1}(\dot{q}_k)^2 -u(q), 
\ee
where $u(q)$ is a potential energy of the system. Then,
Eq. (\ref{5}) becomes
\be \label{Lem} \ddot{q_k}=-\frac{\partial u}{\partial q_k}+
\lambda \frac{\partial f}{\partial \dot{q}_k} ,\quad (k=1,...,n) . \ee
Suppose that the constraint is an integral of motion, i.e., 
${df}/{dt}=0$. Then
\be \label{P1}
\frac{\partial f}{\partial \dot{q}_k} \frac{d\dot{q}_k}{dt}+
\frac{\partial f}{\partial  ( _aD^{\alpha}_t q_k)} 
\frac{d (_aD^{\alpha}_t q_k)}{dt}
+\frac{\partial f}{\partial  ( _tD^{\alpha}_b q_k)} 
\frac{d (_tD^{\alpha}_b q_k)}{dt}
+\frac{\partial f}{\partial q_k} \frac{dq_k}{dt}=0 . \ee
Equation (\ref{P1}) can be presented as
\be \label{CE-t} \frac{\partial f}{\partial \dot{q}_k} \ddot{q_k}
+\frac{\partial f}{\partial  ( _aD^{\alpha}_t q_k)} D^1_t \ _aD^{\alpha}_t q_k
+\frac{\partial f}{\partial  ( _tD^{\alpha}_b q_k)} D^1_t \ _tD^{\alpha}_b q_k
+\frac{\partial f}{\partial q_k} \dot{q}_k=0 . \ee
Substitution of (\ref{Lem}) into (\ref{CE-t}) gives 
\be \frac{\partial f}{\partial \dot{q}_k} 
\left(-\frac{\partial u}{\partial q_k}+
\lambda \frac{\partial f}{\partial \dot{q}_k} \right)
+\frac{\partial f}{\partial ( _aD^{\alpha}_t q_k)} D^1_t \ _aD^{\alpha}_t q_k
+\frac{\partial f}{\partial ( _tD^{\alpha}_b q_k)} D^1_t \ _tD^{\alpha}_b q_k
+\frac{\partial f}{\partial q_k} \dot{q}_k=0 . \ee
From this equation, one can obtain the Lagrange multiplier $\lambda$:
\be \label{P3}
\lambda=
\left(\frac{\partial f}{\partial \dot{q_m}} 
\frac{\partial f}{\partial \dot{q}_m} \right)^{-1}
\left(\frac{\partial f}{\partial \dot{q_l}} 
\frac{\partial u}{\partial q_l}
-\frac{\partial f}{\partial ( _aD^{\alpha}_t q_l)} D^1_t \ _aD^{\alpha}_t q_l
-\frac{\partial f}{\partial ( _tD^{\alpha}_b q_l)} D^1_t \ _tD^{\alpha}_b q_l
-\frac{\partial f}{\partial q_l} \dot{q_l} \right) . \ee
Insertion of Eq. (\ref{P3}) into Eq. (\ref{Lem}) yields 
\be \label{Lem3} \ddot{q_k}=
-\frac{\partial u}{\partial q_k}+
\frac{\partial f}{\partial \dot{q}_k}
\left(\frac{\partial f}{\partial \dot{q_m}}  
\frac{\partial f}{\partial \dot{q_m}} \right)^{-1}
\left(\frac{\partial f}{\partial \dot{q_l}} \frac{\partial u}{\partial q_l}
-\frac{\partial f}{\partial ( _aD^{\alpha}_t q_l)} D^1_t \ _aD^{\alpha}_t q_l
-\frac{\partial f}{\partial ( _tD^{\alpha}_b q_l)} D^1_t \ _tD^{\alpha}_b q_l
-\frac{\partial f}{\partial q_l} \dot{q_l} \right) . \ee
These equations describe holonomic system
that is equivalent to the nonholonomic one with
fractional constraint.
For any motion of the system, we have $f=0$.
If the initial values satisfy the constraint condition 
$f(q(0),\dot{q}(0),\ _aD^{\alpha}_t q(0),\ _tD^{\alpha}_b q(0))=0$,
then the solution of Eq. (\ref{Lem3}) describes a motion
of the  system (\ref{L}) with fractional constraint (\ref{fnhc}).

\subsection{Linear Fractional Constraint}

Suppose that the constraint (\ref{fnhc}) is linear with respect
to integer derivatives $\dot{q}_k$, i.e.,
\be \label{FC-E}  f= 
a_k \dot{q}_k+\beta(\ _aD^{\alpha}_t q,\ _tD^{\alpha}_b q,q) .\ee
In this case, $R_k=a_k$, 
and Eqs. (\ref{Lem3}) can be presented as
\be \label{Lem5} \ddot{q_k}=
-\sum^n_{l=1}\left(\delta_{kl}-\frac{a_ka_l}{a^2}\right)
\frac{\partial u}{\partial q_l}-\frac{a_k}{a^2} 
\sum^n_{l=1} \left( 
\frac{\partial f}{\partial ( _aD^{\alpha}_t q_l)} D^1_t \ _aD^{\alpha}_t q_l
+\frac{\partial f}{\partial ( _tD^{\alpha}_b q_l)} D^1_t \ _tD^{\alpha}_b q_l
+\frac{\partial \beta}{\partial q_l} \dot{q_l} \right) , \ee
where $a^2=\sum^n_{k=1} a_ka_k$. 
If  
\be 
\beta ( _aD^{\alpha}_t q,\ _tD^{\alpha}_b q,q)=b_k \ _aD^{\alpha}_t q_k , 
\ee
then
\be f= a_k \dot{q}_k+ b_k \ _aD^{\alpha}_t q_k .\ee 
This constraint is linear with respect
to integer derivative $\dot{q}_k$ and 
fractional derivatives $ _aD^{\alpha}_t q_k$.
Then the equations of motion are
\be \label{Lem6} \ddot{q_k}=
-\sum^n_{l=1}\left(\delta_{kl}-\frac{a_ka_l}{a^2}\right)
\frac{\partial u}{\partial q_l}-
\sum^n_{l=1}\frac{a_kb_l}{a^2} D^1_t \ _aD^{\alpha}_t q_l . \ee
Using Proposition 1, we obtain
\be \label{Lem6b} 
\ddot{q_k}=-\sum^n_{l=1}\left(\delta_{kl}-\frac{a_ka_l}{a^2}\right)
\frac{\partial u}{\partial q_l}-
\sum^n_{l=1}\frac{a_kb_l}{a^2} 
\left( _aD^{\alpha+1}_t q_l+
\frac{t^{m-\alpha-1}}{\Gamma(m-\alpha)} q^{(m)}_l(a) \right). 
\ee
where $q^{(m)}(a)=(D^m_t q(t))_{t=a}$. 
As the result, we get the fractional equations of motion with
Caputo derivative of order $\alpha+1$.
The nonholonomic systems (\ref{Lem6}) with integer $\alpha$ are 
considered in Refs. \cite{MPLB2003,AP2005-1}.

\subsection{One-dimensional Example}

In one-dimensional case ($n=1$), Eq. (\ref{Lem6}) with $\ _0D^{\alpha}_t q$ 
has the form
\be \label{Lem7} \ddot{q}=-\frac{b_1}{a_1} D^1_t \ _0D^{\alpha}_t q . \ee
Eq. (\ref{Lem7}) can be presented as
\be D^1_t[ \dot{q}+(b_1/a_1)  _0D^{\alpha}_tq ]=0 . \ee
Then 
\be 
\dot{q}+(b_1/a_1)  _0D^{\alpha}_tq =C_0 .
\ee
Supposing $\alpha>1$, and using proposition 1, we get
\be
D^1_t [ q+(b_1/a_1)  _0D^{\alpha-1}_t q] =
\frac{b_1t^{m-\alpha}}{a_1\Gamma(m-\alpha+1)}q(0)+C_0 .
\ee
As the result, we obtain 
\be \label{Eq1}
_0D^{\alpha-1}_t q+(a_1/b_1)q=
\frac{t^{m-\alpha+1}}{\Gamma(m-\alpha+2)}q(0)+C_1 t +C_2 ,
\ee
where we use $ x \Gamma(x)=\Gamma(x+1)$, and $C_1=C_0a_1/b_1$.

For $2<\alpha<3$, Eq. ({\ref{Eq1}}) describes 
the linear fractional oscillator  
\be \label{11}
_0D^{\alpha-1}_t q(t)+\omega^2 q(t)=Q(t) ,
\ee
where $\omega^2=(a_1/b_1)$ is dimensionless "frequency", and
$Q(t)$ is the external force:
\[ 
Q(t)=\frac{t^{m-\alpha+1}}{\Gamma(m-\alpha+2)}q(0)+C_1 t +C_2 . 
\]
The Caputo fractional derivative $ _0D^{\alpha-1}_t$ 
allows us to use the regular initial conditions \cite{Podlubny}
for Eq. (\ref{11}). The linear fractional oscillator  
is an object of numerous investigations 
\cite{GM2,GM3,M,ZSE,TZ2,Stanislavsky,AHC,AHEC,T,RR}
because of different applications.

The exact solution \cite{GM2,GM3} of Eq. (\ref{11}) for $2<\alpha<3$ is 
\be \label{11a}
q(t)=q(0) E_{\alpha-1,1}(-\omega^2 t^{\alpha-1})+
t q^{\prime}(0) E_{\alpha-1,2}(-\omega^2 t^{\alpha-1}) -
\int^t_0 Q(t-\tau) \dot{q}_{0}(\tau) d \tau,
\ee
where 
\be \label{11b}
E_{\alpha,\beta}(z)=\sum^{\infty}_{k=0} 
\frac{z^k}{\Gamma(\alpha k+\beta)} 
\ee
is the generalized two-parameter Mittag-Leffler function \cite{ML1,ML2}, and 
\[ q_0(\tau)=E_{\alpha-1,1}(-\omega^2 \tau^{\alpha-1}) . \]
The decomposition of (\ref{11a}) is \cite{GM2}:
\be \label{ES1}
q(t)=q(0)\left[f_{\alpha,0}(t)+g_{\alpha,0}(t)\right]+
t \dot{q}(0)\left[f_{\alpha,1}(t)+g_{\alpha,1}(t)\right] -
\int^t_0 Q(t-\tau) \dot{q}_{0}(\tau) d \tau,
\ee
where
\[ f_{\alpha,k}(t)=\frac{(-1)^k}{\pi} \int^{\infty}_0 e^{-rt}
\frac{r^{\alpha-1-k} \sin(\pi \alpha)}{
r^{2\alpha}+2r^{\alpha} \cos(\pi \alpha)+1} dr,
\]
\be \label{ES5}
g_{\alpha,k}(t)=\frac{2}{\alpha} e^{t \cos(\pi / \alpha)} \
\cos \left[ t \sin(\pi/\alpha)-{\pi k}/{\alpha } \right],
\quad (k=0,1) .
\ee
For the initial conditions $q(0)=1$, and $\dot{q}(0)=0$:
\be \label{ES6}
q(t)=E_{\alpha} (-t^{\alpha})=
f_{\alpha,0}(t)+g_{\alpha,0}(t)- 
\int^t_0 Q(t-\tau) [\dot{f}_{\alpha,0}(\tau)+\dot{g}_{\alpha,0}(\tau)] d \tau.
\ee
The first term in (\ref{ES6}) decay in power law with time while
the second term decays exponentially \cite{GM2,GM3,ZSE,TZ2}.

\subsection{Two-dimensional Examples}

In the two-dimensional case ($n=2$), Eq. (\ref{Lem6}) has the form
\be \label{Lem8a} \ddot{q_1}=
-\frac{a^2_2}{a^2_1+a^2_2}\frac{\partial u}{\partial q_1}
+\frac{a_1a_2}{a^2_1+a^2_2}\frac{\partial u}{\partial q_2}
-\frac{a_1b_1}{a^2_1+a^2_2} D^1_t \ _aD^{\alpha}_t q_1
-\frac{a_1b_2}{a^2_1+a^2_2} D^1_t \ _aD^{\alpha}_t q_2 ; \ee
\be \label{Lem8b} \ddot{q_2}=
-\frac{a^2_1}{a^2_1+a^2_2}\frac{\partial u}{\partial q_2}
+\frac{a_1a_2}{a^2_1+a^2_2}\frac{\partial u}{\partial q_1}
-\frac{a_2b_1}{a^2_1+a^2_2}D^1_t \ _aD^{\alpha}_t  q_1
-\frac{a_2b_2}{a^2_1+a^2_2}D^1_t \ _aD^{\alpha}_t  q_2 . \ee

Let us consider the special cases of these equations.\\

\noindent
(1) Suppose $a_1=0$; then 
\be \label{Lem9} \ddot{q_1}=
-\frac{\partial u}{\partial q_1}; \quad
\ddot{q_2}=-\frac{b_1}{a_2} D^1_t \ _aD^{\alpha}_t q_1
-\frac{b_2}{a_2} D^1_t \ _aD^{\alpha}_t q_2 . \ee
If $a_1=0$, and $b_2=0$, then (\ref{Lem9}) are 
\be \label{Lem9c} \ddot{q_1}=
-\frac{\partial u}{\partial q_1}, \quad
\ddot{q_2}= -\frac{b_1}{a_2} D^1_t \ _aD^{\alpha}_t  q_1 . \ee

\vskip 3mm

\noindent
(2) Suppose $b_1=0$ and $a_1=a_2=c$; then we have
\be \label{Lem10a} \ddot{q_1}=
-\frac{1}{2}\frac{\partial u}{\partial q_1}
+\frac{1}{2}\frac{\partial u}{\partial q_2}
-\frac{b_2}{2c} D^1_t \ _aD^{\alpha}_t q_2 ; \ee
\be \label{Lem10b} \ddot{q_2}=
-\frac{1}{2}\frac{\partial u}{\partial q_2}
+\frac{1}{2}\frac{\partial u}{\partial q_1}
-\frac{b_2}{2c} D^1_t \ _aD^{\alpha}_t q_2 . \ee
Using 
\[ x=\frac{q_1+q_2}{2}, \quad y=\frac{q_1-q_2}{2} ,
\quad g=b_2 /c , \]
we can rewrite Eqs. (\ref{Lem10a}) and (\ref{Lem10b})
in the form 
\be \label{53}
\ddot{x}=-g D^1_t \ _aD^{\alpha}_tx+ g D^1_t \ _aD^{\alpha}_t  y, 
\quad
\ddot{y}=-\frac{\partial U}{\partial y} ,
\ee
where $U(x,y)=u(q_1,q_2)=u(x+y,x-y)$.
If $U= K(x) y+s(x)$, then Eq. (\ref{53}) is
\be \label{fs}
\ddot{x}=-g D^1_t \ _aD^{\alpha}_t x + g D^1_t \ _aD^{\alpha}_t y, 
\quad \ddot{y}=-K(x) .
\ee
Using $D^{\alpha}_t y=D^{\alpha-2}_t \ddot{y}$, 
Eq. (\ref{fs}) gives
\be \label{dd} 
\ddot{x}=-g D^1_t \ _aD^{\alpha}_t x-g D^1_t \ _aD^{\alpha-2}_t  K(x). 
\ee
Then
\[ 
D^1_t[ \dot{x}+g \ _aD^{\alpha}_t x+g \ _aD^{\alpha-2}_t  K(x)]=0 .
\]
For $\alpha >2 $, then we can get 
\be \label{FGL} \dot{x}+g^{-1} _aD^{\alpha}_t x +
g _aD^{\alpha-2}_t K(x)+C=0,  \ee
where $C$ is a constant.
Using $ _aD^{\varepsilon}_t \ _aJ^{\varepsilon}_t f(t)=f(t)$,
and $ _aD^{\varepsilon}_t \ _aD^{\alpha}_t f(t)=D^m_t f(t)$,
where $\varepsilon=m-\alpha$, Eq. (\ref{FGL})
can be written as
\be \label{a-2}
D^{\varepsilon} x+g^{-1} x^{(m)} + g _a D^{m-2}_t K(x)=0.
\ee
If $m=2$ ($1<\alpha<2$), then
\be \label{a-3}
g^{-1} \ddot{x} + D^{\varepsilon} x+g _a K(x)=0.
\ee
This equation can be considered as an equation of nonlinear fractional
oscillator \cite{ZSE,TZ2}, where fractional derivative describes 
the power dumping.

\section{Fractional Conditional Extremum}

\subsection{Extremum for Fractional Constraint}

Let us consider the stationary value of an action integral
\[ \delta \int^{b}_{a} dt \ L(q,\dot{q})=0 , \]
for the lines that satisfy the constraint equation $f(q,\dot{q})=0$. 
Using the Lagrange multiplier $\mu=\mu(t)$, we get a variational equation
\[ \delta \int^{b}_{a} dt 
\left[ L(q,\dot{q})+\mu f(q,\dot{q}) \right]=0 . \]
Then the Euler-Lagrange equations \cite{Rum} are
\be \label{Ru}
\frac{d}{dt} \frac{\partial L}{\partial \dot{q}_k}-
\frac{\partial L}{\partial q_k}=
\mu \left(\frac{\partial f}{\partial q_k}-
\frac{d}{dt} \frac{\partial f}{\partial \dot{q}_k} \right)
-\dot{\mu} \frac{\partial f}{\partial \dot{q}_k},
\quad (k=1,...,n) . \ee
Note that these equations consist of the derivative 
of Lagrange multiplier $\dot{\mu}$. The proof of Eq. (\ref{Ru})
is realized in \cite{Rum}.

For the fractional constraint
\be \label{66}
f(q,\dot{q},\ _aD^{\alpha}_t q,\ _tD^{\alpha}_b q)=0,  \ee
we can define the Lagrangian as
\be \label{cal-L} 
{\cal L}(q,\dot{q},\ _aD^{\alpha}_t q,\ _tD^{\alpha}_b q,\lambda) =
L(q,\dot{q})+\mu(t) f(q,\dot{q},\ _aD^{\alpha}_t q,\ _tD^{\alpha}_b q).
\ee
Using the Agrawal variational equation \cite{Agrawal}, we obtain the
Euler-Lagrange equations
\be \label{Ham0} \frac{\partial {\cal L}}{\partial q_k}
-\frac{d}{dt} \frac{\partial {\cal L}}{\partial \dot{q}_k}
+ _aD^{\alpha}_t \frac{\partial {\cal L}}{\partial ( _aD^{\alpha}_t q_k)}
+ _tD^{\alpha}_b \frac{\partial {\cal L}}{\partial ( _tD^{\alpha}_b q_k)}
=0, \quad (k=1,...,n) .\ee
Substitution of Eq. (\ref{cal-L}) into Eq. (\ref{Ham0}) gives
\be \label{33} 
\frac{\partial L}{\partial q_k}
-\frac{d}{dt} \frac{\partial L}{\partial \dot{q}_k}
+ \mu \frac{\partial f}{\partial q_k}
+\ _aD^{\alpha}_t \left( \mu \frac{\partial f}{\partial ( _aD^{\alpha}_t q_k)} 
\right)
+\ _tD^{\alpha}_b \left( \mu \frac{\partial f}{\partial ( _tD^{\alpha}_b q_k)} 
\right)
-\frac{d}{dt} \left( \mu \frac{\partial f}{\partial \dot{q}_k}\right)=0 .\ee
These equations describe the fractional conditional extremum.

Let us consider applicability 
of stationary action principle for mechanical systems 
with fractional nonholonomic constraints.
The equations of motion are derived from 
the d'Alembert-Lagrange principle.
The fractional conditional extremum can be obtained from 
stationary action principle. 
In general, these equations are not equivalent \cite{Rum}. 
The condition of this equivalence for fractional constraints 
is suggested in the proposition. \\

{\bf Proposition 2.}  
{\it Equations (\ref{5}) and (\ref{33}) 
for nonholonomic system with fractional constraint (\ref{66})
have the equivalent set of solutions if the conditions  
\be \label{cond}
\left[ \mu \frac{\partial f}{\partial q_k}
+\ _aD^{\alpha}_t \left( \mu \frac{\partial f}{\partial _aD^{\alpha}_t q_k} 
\right)
+\ _tD^{\alpha}_b \left( \mu \frac{\partial f}{\partial _tD^{\alpha}_b q_k} 
\right)
-\frac{d}{dt} \left( \mu \frac{\partial f}{\partial \dot{q}_k}\right) 
\right] \delta q_k=0 , \quad
\frac{\partial f}{\partial \dot{q}_k} \delta q_k=0 .\ee
are satisfied.} \\

{\bf Proof}. 
To prove the proposition, we multiply Eqs. (\ref{5}) and (\ref{33}) 
on the variation $\delta q^k$ and consider a sum with respect to $k$:
\be \label{5c} \left( \frac{d}{dt} \frac{\partial L}{\partial \dot{q}_k} -
\frac{\partial L}{\partial q_k}\right) \delta q_k=
\lambda \frac{\partial f}{\partial \dot{q}_k} \delta q_k, \ee

\be \label{33c} 
\left( \frac{\partial L}{\partial q_k}-
\frac{d}{dt} \frac{\partial L}{\partial \dot{q}_k} \right) \delta q_k
+ \left[ \mu \frac{\partial f}{\partial q_k}
+\ _aD^{\alpha}_t\left( \mu \frac{\partial f}{\partial ( _aD^{\alpha}_t q_k)} 
\right)
+\ _tD^{\alpha}_b\left( \mu \frac{\partial f}{\partial ( _tD^{\alpha}_b q_k)} 
\right)
-\frac{d}{dt}\left( \mu \frac{\partial f}{\partial \dot{q}_k}\right) 
\right] \delta q_k =0 ,\ee
From the definition of variations (\ref{Ch}), Eq. (\ref{5c}) is
\be \label{5d} \left( \frac{d}{dt} \frac{\partial L}{\partial \dot{q}_k} -
\frac{\partial L}{\partial q_k}\right) \delta q_k=0 , \ee
Substituting Eq. (\ref{5d}) into Eq. (\ref{33c}), we obtain (\ref{cond}).\\

It is known \cite{Rum}, that stationary action principle 
cannot be derived from the d'Alembert-Lagrange principle
for wide class of nonholonomic and non-Hamiltonian systems.
The same can be applied to the case of nonlinear 
fractional nonholonomic constraints.

\subsection{Hamilton's Approach}

Using Eq. (\ref{Ham0}), we define the momenta
\be \label{mom1}
p_k=\frac{\partial {\cal L}}{\partial \dot{q}_k}=
\frac{\partial L}{\partial \dot{q}_k}+\mu
\frac{\partial f}{\partial \dot{q}_k} ,
\ee
and the Hamiltonian 
\be {\cal H}(q,p)=p_k\dot{q}-{\cal L}, \ee
where ${\cal L}=L+\mu f$.
Eq. (\ref{Ham0}) gives
\be 
\frac{dp_k}{dt}= \frac{\partial {\cal H}}{\partial q_k}
+\ _aD^{\alpha}_t 
\left( \mu \frac{\partial f}{\partial ( _aD^{\alpha}_t q_k)} 
\right)
+\ _tD^{\alpha}_b 
\left( \mu \frac{\partial f}{\partial ( _tD^{\alpha}_b q_k)} 
\right).
\ee
To simplify our calculations, we consider the Lagrangian 
\[ L=\frac{1}{2}(\dot{q})^2-u(q) , \]
and the fractional nonholonomic constraint
\be \label{con}
f=A_k(q,\ _aD^{\alpha}_t q) \ \dot{q}_k=0, \quad (\alpha \not=1). 
\ee
From Eqs. (\ref{mom1}) and (\ref{con}), we obtain
\be \label{mom2}
p_k=\dot{q}_k+\mu A_k(q,\ _aD^{\alpha}_t q).
\ee 
Then the Hamilton equations are
\be \label{dq1}
\dot{q}_k=p_k-\mu A_k(q,\ _aD^{\alpha}_t q),
\ee 
\be \label{dp1}
\dot{p}_k=-\frac{\partial u(q)}{\partial q_k}+\mu(t)\dot{q}_l 
\frac{\partial A_l}{\partial q_k}+
\ _aD^{\alpha}_t \left(\mu \dot{q}_l 
\frac{\partial A_l(\ _aD^{\alpha}_t q)}{\partial _aD^{\alpha}_t q_k} \right).
\ee
To find the Lagrange multiplier $\mu=\mu(t)$, we 
multiply Eq. (\ref{mom2}) on the functions $a_k$ and 
consider the sum with respect to $k$:
\be \label{80}
A_kp_k=A_k \dot{q}_k+\mu A_kA_k=\mu A^2.
\ee
Here, we use the constraint (\ref{con}), and the notation 
$A^2=A_kA_k$, where $A_k=A_k(q,\ _aD^{\alpha}_t q)$. 
From (\ref{80}), we get
\be \label{mu}
\mu=\frac{A_kp_k}{A^2} .
\ee 
Substitution of (\ref{mu}) into Eqs. (\ref{dq1}) and (\ref{dp1}) gives 
\be \label{dq}
\dot{q}_k=\left( \delta_{kl}-\frac{A_kA_l}{A^2} \right) p_l,
\ee 
\be \label{dp}
\dot{p}_k=-\frac{\partial u(q)}{\partial q_k}+
\frac{A_mp_m}{A^2} \frac{\partial A_l}{\partial q_k}\dot{q}_l+
\ _aD^{\alpha}_t \left( \frac{A_m p_m}{A^2} 
\frac{\partial A_l}{\partial _aD^{\alpha}_t q_k} \dot{q}_l \right).
\ee
If $a_k=0$, then we have usual equations of motion for
Hamiltonian systems.
Note that we derive Hamiltonian equations from Euler-Lagrange
equations without using the Legendre transformation,
which is typically used.

\section{Conclusion}

The classical mechanics of nonholonomic systems has recently 
been employed to study a wide variety of problems
in the molecular dynamics \cite{FS}. 
In molecular dynamics calculations, nonholonomic systems 
can be exploited to generate statistical ensembles as 
the canonical, isothermal-isobaric and isokinetic ensembles 
\cite{EHFML,HG,EM,Nose,GA,Tuck1,Tuck2,Tuck3,Ram,MPLB2003,AP2005-1}.
Using fractional nonholonomic constraints, we can consider a
fractional extension of the statistical mechanics of 
conservative Hamiltonian systems to a much broader class of systems.
Let us point out some nonholonomic systems 
that can be generalized by using the nonholonomic constraint with
fractional derivatives.

(1) In the papers \cite{EHFML,HG,EM,Nose}, the constant 
temperature systems with minimal Gaussian constraint are considered.
These systems are non-Hamiltonian ones and they
are described by the non-potential forces that are proportional to 
the velocity, and the Gaussian nonholonomic constraint. 
Note that this constraint can be represented 
as an addition term to the non-potential force \cite{AP2005-1}.

(2) In the papers \cite{MPLB2003,AP2005-1}, 
the canonical distribution is considered as a stationary solution 
of the Liouville equation for a wide  class of non-Hamiltonian system.
This class is defined by a very simple condition:
the power of the non-potential forces must be proportional
to the velocity of the Gibbs phase (elementary phase volume) change.
This condition defines the general constant temperature systems.
Note that the condition is a nonholonomic constraint. This constraint 
leads to the canonical distribution as a stationary solution 
of the Liouville equations.
For the linear friction, we derived the constant temperature systems. 
A general form of the non-potential forces 
is derived in Ref. \cite{AP2005-1}.


\section*{Acknowledgments}

This work was supported by the Office of Naval Research,
Grant No. N00014-02-1-0056, the U.S. Department
of Energy Grant No. DE-FG02-92ER54184, and the NSF
Grant No. DMS-0417800. 
VET thanks the Courant Institute of Mathematical Sciences
for support and kind hospitality.


\end{document}